\documentclass[10pt,twocolumn,twoside]{IEEEtran}

\usepackage{amsfonts}
\usepackage{amssymb}
\usepackage{cite}
\usepackage{graphicx}
\usepackage[cmex10]{amsmath}
\usepackage[]{algorithm}
\usepackage{algorithmic}
\usepackage{subeqnarray}
\usepackage{cases}
\usepackage{cleveref}
\usepackage{comment}
\usepackage{subfigure}
\IEEEoverridecommandlockouts

\ifCLASSINFOpdf
\else
\fi

\begin{document}

%
\title{On the Energy Self-Sustainability of IoT via Distributed Compressed Sensing}

%
%

\author{
    \authorblockN{Wei Chen\authorrefmark{1}, Nikos Deligiannis\authorrefmark{2}\authorrefmark{5}
        Yiannis Andreopoulos\authorrefmark{3},
        Ian J. Wassell\authorrefmark{4}\\
    }
    \authorblockA{
        \authorrefmark{1}
        State Key Laboratory of Rail Traffic Control and Safety, Beijing Jiaotong University, China}\\
    \authorblockA{
        \authorrefmark{2}
        Electronics and Informatics department, Vrije Universiteit Brussel, Belgium}\\
    \authorblockA{
        \authorrefmark{3}
        Department of Electronic \& Electrical Engineering, University College London, UK}\\
    \authorblockA{
        \authorrefmark{4}
        Computer Laboratory, University of Cambridge, UK}\\
    \authorblockA{
        \authorrefmark{5}
        imec, Belgium}
}

\maketitle

\begin{abstract}
This paper advocates the use of the distributed compressed sensing (DCS) paradigm to deploy energy harvesting (EH) Internet of Thing (IoT) devices for energy self-sustainability. We consider networks with signal/energy models that capture the fact that both the collected signals and the harvested energy of different devices can exhibit correlation. We provide theoretical analysis on the performance of both the classical compressive sensing (CS) approach and the proposed distributed CS (DCS)-based approach to data acquisition for EH IoT. Moreover, we perform an in-depth comparison of the proposed DCS-based approach against the distributed source coding (DSC) system. These performance characterizations and comparisons embody the effect of various system phenomena and parameters including signal correlation, EH correlation, network size, and energy availability level. Our results unveil that, the proposed approach offers significant increase in data gathering capability with respect to the CS-based approach, and offers a substantial reduction of the mean-squared error distortion with respect to the DSC system.
\end{abstract}

\begin{IEEEkeywords}
Distributed compressed sensing, energy harvesting, internet of things, energy self-sustainability.
\end{IEEEkeywords}

%
\IEEEpeerreviewmaketitle

\section{Introduction}
\IEEEPARstart{D}{evices} with energy self-sustainability (ESS) are desired for the Internet of Things (IoT) and 6G, i.e. 6th generation of mobile communications~\cite{8820755}. To achieve ESS, future communication devices are expected to be equipped with energy harvesters (e.g., piezoelectric, thermoelectric and photovoltaic) to substantially increase their autonomy and lifetime~\cite{8664000,8938189,8913774}. The use of energy harvesting (EH) has been emerging in various IoT applications, e.g., greenhouse monitoring using solar energy and super capacitor storage, remote sensing of wind-driven wildfire spread, and radio frequency EH in structural health monitoring network. However, it is also recognized that the gap between EH supply and the devices' energy demand is not likely to close in the near future due to the surge in demand for more data-intensive applications.

\par
These considerations have motivated the design of energy efficient data sensing and coding schemes~\cite{7001194,8777159}. Such approaches rely on the intra-sensor data correlation but fail to exploit the correlations amongst data captured by different devices. Rooted in the theoretical results of Slepian and Wolf \cite{slepianwolf} and Wyner and Ziv \cite{wynerziv}, distributed source coding (DSC) schemes exploit inter-sensor data correlation via joint decoding~\cite{7083694}. While offering low-complexity solutions, the performance of DSC systems is highly dependent on knowledge of the correlation statistics, and extending DSC to the multiterminal case is a challenging problem in practice.

\par
Compressive sensing (CS) is a sampling paradigm that can reduce energy consumption associated with data acquisition and transmission~\cite{5595724,6168432,6548096,7390294,8480642}. By exploiting the CS principle, the scheme in~\cite{5595724} shows that a reduced number of weighted sums of sensor readings (instead of individual readings) can be delivered to the collection unit, thereby reducing both communication and computation costs. Alternatively, in~\cite{6168432}, an adaptive and nonuniform compressive sampling approach is applied to improve the energy efficiency of devices. Moreover, unbalanced costs of different devices are considered for scheduled sensing to prolong the system lifetime in~\cite{7390294}. In~\cite{8480642}, a CS-based prejudiced random sensing strategy is proposed to attain a desired tradeoff between the overall energy consumption and the sensing accuracy. Finally, the CS principles have been extended to the multiterminal case by means of distributed compressed sensing (DCS)~\cite{baron2006distributed,6502243,8532355}, which exploits both intra- and inter-sensor data correlations via joint reconstruction at the collection unit.

\par
Regarding energy-efficient data transmission, existing works focus on the design of an intelligent point-to-point wireless communication system with EH capability, or network-level energy management with multiple IoT devices and base stations (BSs). Yet, these solutions do not explicitly integrate two fundamental mechanisms: \emph{energy diversity} and \emph{sensing diversity}.

\par
This paper advocates the use of DCS for ESS in IoT applications. The key attributes of the proposed approach that lead to ESS are as the following: Due to signal correlations, the number of measurements at the various devices can be substantially lower than the data dimensionality without compromising data recovery. In addition, a variable number of measurements can be allocated to different devices (subject to EH constraints) without compromising data recovery. Hence, we argue that, due to the energy diversity (associated with the EH process) and the sensing diversity (associated
with the DCS process), we can match the energy supply to the energy demand. In this way, we can unlock the possibility for ESS. Our contributions lead to ESS are as follows:
\begin{itemize}
  \item We propose a DCS-based sensing approach to unlock ESS in EH IoT networks by matching the energy demand to the profile of energy supply. Our approach is fundamentally different from other CS or DCS approaches that focus purely on the reduction in the required number of measurements;
  \item We derive a lower bound to the probability of incorrect data reconstruction (PIDR) for both a CS-based data acquisition scheme, which only exploits intra-sensor correlations, and the DCS-based data acquisition scheme, which exploits both intra- and inter-sensor correlations;
  \item We analyze the performance of the proposed approach via numerical simulations that embody the effect of various system phenomena and parameters (such as signal correlation, energy harvesting correlation, network size, and energy availability level). In particular, we show that there exist an optimal number of signals for joint reconstruction;
  \item We conduct an in-depth experimental comparison of the proposed DCS system against the DSC approach using real sensor data and we demonstrate the superiority of our solution. To the best of our knowledge, this is the first time where DCS\ and DSC are compared against each other in a systematic manner.
\end{itemize}


\par
The following notational conventions are adopted throughout the paper. Lower-case letters denote
scalars; boldface upper-case letters denote matrices; boldface lower-case letters denote column
vectors; calligraphic upper-case letters denote support sets and $\mathbf{0}$
denotes a vector or a matrix with all zeros. The superscript $(\cdot)^T$ denotes
matrix transpose. The $\ell_0$ norm, the $\ell_1$ norm, and the
$\ell_2$ norm of vectors, are denoted by $\|\cdot\|_{0}$, $\|\cdot\|_{1}$, and $\|\cdot\|_{2}$,
respectively. $\text{Pr}(\cdot)$ and $\emph{P}_{x}(\cdot)$ denote the probability and the probability density function (PDF) of $x$ respectively.

\begin{figure}[t]%
\centering%
\includegraphics[width=0.30\textwidth]{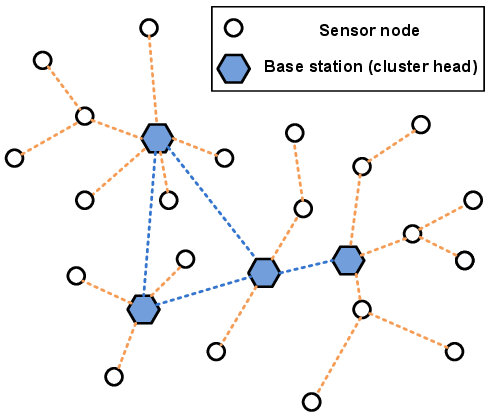}%
\DeclareGraphicsExtensions. \caption{A typical cluster-based IoT architecture.}
\label{fig:wsn-model}
\end{figure}%
\begin{figure}[t]%
\centering%
\includegraphics[width=0.35\textwidth]{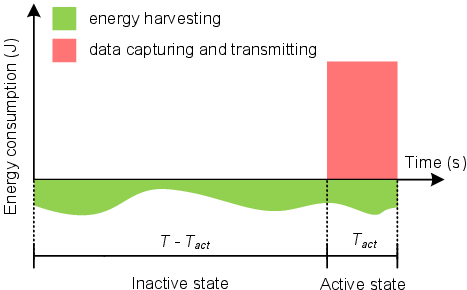}%
\DeclareGraphicsExtensions. \caption{Typical energy consumption profile of a data acquisition and
EH scheme.} \label{fig:energy-harvesting}
\end{figure}%

\section{System Description}
We consider a typical cluster-based IoT architecture, where a set of devices periodically conveys data to one or more base stations (BSs) that form the aggregation point of the cluster (see Fig.~\ref{fig:wsn-model}). We assume slotted transmission such that within a time slot of $T$ seconds the devices are active for $T_\text{act}$ seconds in order to capture and transmit data and are inactive for $T-T_\text{act}$ seconds. Energy may be harvested from the environment during each time slot $T$ and can be stored in a battery, as shown in Fig.~\ref{fig:energy-harvesting}. We assume that, upon activation, the devices converge into a balanced time-frequency steady-state mode where each device is associated with a BS using a particular channel (or joins a synchronized channel hopping schedule) in order to convey data without collisions. We also assume fading, external interference and other non-idealities in packet transmissions are dealt with via the physical-layer modulation and coding mechanisms of standards such as IEEE 802.15.4. Therefore, without loss of generality, from the sensing and processing side, data transmission is taken to be a lossless process with any non-idealities dealt with via the lower layers of the protocol stack~\cite{6612900}.

\par
We also consider a data gathering and reconstruction process---which is key to match the energy demand to the energy supply---based on three
steps: \textit{(i)} DCS based data acquisition at the devices, \textit{(ii)} data transmission from the devices to the BS, and \textit{(iii)} DCS based data reconstruction at the BS. These processes, together with the energy consumption model and the EH model, are described in the sequel. Note that the idea and results are presented exclusively for a centralized IoT architecture consisting of $K$ devices that are attached to a single BS. However, our scheme can be straight forwardly generalized to architectures with devices that are attached to multiple BSs, as in Fig. \ref{fig:wsn-model}.

\subsection{DCS Based Data Acquisition and Transmission}
The devices capture low-dimensional projections of the original high-dimensional data during each
activation time $iT-T_\text{act}\leq t\leq iT$, which are given by:
\begin{equation}\label{eq:system_model1}
\mathbf{y}_k(i)=\mathbf{\Phi}_k(i) \mathbf{f}_k(i),
\end{equation}
where $\mathbf{y}_k(i)\in \mathbb{R}^{m_k(i)}$ is the projections vector at the $k$th device
corresponding to the $i$th time interval\footnote{Note that the dimensionality of the projections can vary in different activation times and different devices.}, $\mathbf{f}_k(i)\in \mathbb{R}^{n(i)}$ is the original (Nyquist-sampled) data vector at the $k$th device corresponding to the $i$th time interval, and $\mathbf{\Phi}_k(i) \in \mathbb{R}^{m_k(i)\times n(i)}$ is the projections matrix where $m_k(i)\ll n(i)$ for any time interval $i$ and device $k$. In practice, one may obtain the projections vector from the original data signal using analogue CS encoders~\cite{5948420}, whereby the projections vector is obtained directly from the analogue continuous-time data, or using digital CS encoders~\cite{6155205}, whereby the projections vector is obtained from the Nyquist sampled discrete-time data via (\ref{eq:system_model1}). The devices then convey the low-dimensional projections of the original high-dimensional data to the BS. 

\subsection{DCS Based Data Reconstruction}
We take the signals $\mathbf{f}_k(i)\in \mathbb{R}^{n(i)}$ to admit a sparse representation $\mathbf{x}_k(i)\in \mathbb{R}^{n(i)}$ in some orthonormal basis $\mathbf{\Psi}(i) \in \mathbb{R}^{n(i)\times n(i)}$, i.e.,
\begin{equation}\label{eq:system_model2}
\mathbf{f}_k(i)=\mathbf{\Psi}(i) \mathbf{x}_k(i),
\end{equation}
where $\|\mathbf{x}_k(i)\|_0=s(i)\ll m_k(i) \ll n(i)$.
In addition, we take the sparse representations to obey the sparse common component and innovations (SCCI) model that has been frequently used to capture intra- and inter-signal correlation typical of physical signals (e.g., temperature, humidity) in~\cite{baron2006distributed,6502243}, i.e., we write
\begin{equation}\label{eq:SCCI_model}
\mathbf{x}_k(i)=\mathbf{z}_c(i)+\mathbf{z}_k(i),
\end{equation}
where $\mathbf{z}_c(i)\in \mathbb{R}^{n(i)}$ with $\|\mathbf{z}_c(i)\|_0=s_c'(i)\ll n(i)$ denotes the common component of the sparse representation ${\bf x}_k(i)\in \mathbb{R}^{n(i)}$, which is common to the signals captured by the various devices, and $\mathbf{z}_k(i)\in \mathbb{R}^{n(i)}$ with $\|\mathbf{z}_k(i)\|_0=s'(i)\ll n(i)$ denotes the innovations component of the sparse representation ${\bf x}_k(i)\in \mathbb{R}^{n(i)}$, which is specific to the signals captured by each device. This model applies to scenarios of monitoring specific physical phenomena such as temperature or humidity where the common component models global factors, e.g., the sun and prevailing winds, and the innovations component models local factors, e.g., the terrain and shade. Note that $s_c'(i)+s'(i)\geq s(i)$. Note also that the signal sparsities $s_c'(i)$, $s'(i)$ and $s(i)$, the signal dimensionality $n(i)$, and the orthonormal dictionary $\mathbf{\Psi}(i)$ are in general independent of the activation interval $i$.

\par
In view of the signal model in (\ref{eq:system_model2}) and (\ref{eq:SCCI_model}), it is possible to reconstruct the original signal from the signal projections using either standard CS recovery algorithms or DCS recovery algorithms. CS recovery only considers intra-signal correlation; in contrast, DCS considers both inter- and intra-signal correlation~\cite{baron2006distributed}.

\subsubsection{CS Reconstruction Algorithms}
CS signal reconstruction only assumes that the signals admit a sparse representation in some orthonormal basis, e.g., the discrete Fourier basis and wavelet basis. Therefore, the typical signal reconstruction process behind conventional CS approaches involves solving the following optimization problem to recover individually the original signals captured by the various devices in each activation interval:
\begin{equation}\label{eq:CS_L1}
\begin{split}
\min_{\mathbf{x}_k(i)} \qquad &\|\mathbf{x}_k(i)\|_{1}\\
\text{s.t.} \qquad &\mathbf{A}_k(i)\mathbf{x}_k(i)=\mathbf{y}_k(i),
\end{split}
\end{equation}
where $\mathbf{A}_k(i)=\mathbf{\Phi}_k(i)\mathbf{\Psi}(i) \in \mathbb{R}^{m_k(i)\times n(i)}$. The major practical algorithms for sparse signal reconstruction are surveyed in~\cite{5456165}. Instead of directly dealing with the above convex optimization problem, there are various algorithms and extensions based on sparse Bayesian learning~\cite{1315936,7558157,8453881,8861380}. Nonconvex algorithms for sparse reconstruction is given in~\cite{8332502}. In addition, the rapid development of deep learning (DL) provides a fresh perspective for solving the linear inverse problem. Interested readers may refer to~\cite{BAI2020107729} for a more detailed review of DL based algorithms for linear inverse problems.


\subsubsection{DCS Reconstruction Algorithms}
The signal reconstruction process behind the adopted DCS approach---which exploits the SCCI model in (\ref{eq:system_model2}) together with (\ref{eq:SCCI_model})---involves solving the following optimization problem to recover jointly the original signals captured by various devices in each activation interval~\cite{baron2006distributed}:
\begin{equation}\label{eq:CS_DCS}
\begin{split}
\min_{\tilde{\mathbf{z}}(i)} \qquad &\|\tilde{\mathbf{z}}(i)\|_1\\
\text{s.t.} \qquad &\tilde{\mathbf{A}}(i)\tilde{\mathbf{z}}(i)=\tilde{\mathbf{y}}(i),
\end{split}
\end{equation}
where $\tilde{\mathbf{z}}(i)=\left[\mathbf{z}_c(i)^T\ \mathbf{z}_1(i)^T\ \ldots\
\mathbf{z}_K(i)^T\right]^T\in \mathbb{R}^{(K+1)n(i)}$ is the extended sparse signal vector,
$\tilde{\mathbf{y}}(i)=\left[\mathbf{y}_1(i)^T\ \ldots\ \mathbf{y}_K(i)^T\right]^T\in
\mathbb{R}^{\sum_{k=1}^K m_k(i)}$ is the extended measurements vector, and
$\tilde{\mathbf{A}}(i)\in \mathbb{R}^{\left(\sum_{k=1}^K m_k(i)\right) \times (K+1)n(i)}$ is the
extended sensing matrix given by
\[\tilde{\mathbf{A}}(i)=
\begin{bmatrix}
\mathbf{A}_1(i)&\mathbf{A}_1(i)&\mathbf{0}&\mathbf{0}  & \cdots & \mathbf{0}      \\
\mathbf{A}_2(i)&\mathbf{0}&\mathbf{A}_2(i)&\mathbf{0}  & \cdots & \mathbf{0}      \\
\vdots &  &    &                    & \ddots     & \vdots \\
\mathbf{A}_K(i)&\mathbf{0}&\mathbf{0}&\mathbf{0}    & \cdots & \mathbf{A}_K(i)
\end{bmatrix}.\]

\subsection{Energy Consumption and Harvesting Models}
We assume that the devices use all the available energy in their local battery during each activation interval, which is given by:
\begin{equation}\label{eq:energy_budget}\nonumber
\begin{split}
\xi_k^C(i)=\xi_k^H(i),
\end{split}
\end{equation}
where $\xi_k^H(i)$ is the energy harvested by device $k$ in the interval $(i-1)T\leq t< iT$, and $\xi_k^C(i)$ is the
energy consumed by device $k$ in the interval $iT-T_\text{act}\leq t< iT$.


\subsubsection{Energy Consumption Model}
We assume that the energy consumed for sensing, computing and transmitting one measurement (projection) is essentially a constant
$\tau>0$. Hence, the energy consumed by device $k$ during activation interval $i$ is modelled as follows: 
\begin{equation}\label{eq:measurement_energy_relationship}\nonumber
\xi_k^C(i) = \tau m_k(i).
\end{equation}

\subsubsection{Energy Harvesting Model}
We also assume that the energy harvested by the various devices exhibits some degree of correlation. In particular, the energy harvested by device $k$ during activation interval $i$ is modelled as follows:
\begin{equation}\label{eq:harvesting_model}\nonumber
\xi_k^H(i) = \hat{\xi}_c^H(i) + \hat{\xi}_k^H(i),
\end{equation}
where $\hat{\xi}^H_c(i)$ denotes a component of the harvested energy that is common to all devices and $\hat{\xi}^H_k(i)$ denotes a component of the harvested energy that is specific to the $k$th device. We assume that (1) $\hat{\xi}^H_c(i)$ follows an exponential distribution with parameter $\lambda_c>0$ and that $\hat{\xi}^H_k(i), k=1,\ldots,K$ follows an exponential distribution with parameter $\lambda_k>0, k=1,\ldots,K$ \footnote{We assume that $\sum_{k=1}^{K} \lambda_k \neq \lambda_c$. This mathematical technicality does not result in a substantial loss of generality, but is required in order to simplify the ensuing analysis. We would also like to clarify that the use of the DCS paradigm to deploy EH IoT devices for energy self-sustainability is not relying on the distribution assumption, although it helps in developing the theoretical analysis.}~\cite{7500416}; (2) $\hat{\xi}^H_c(i)$ and $\hat{\xi}^H_k(i), k=1,\ldots,K$ are independent; and (3) EH across time slots is independent.

\par
It is clear that this correlated EH model is akin to the signal correlation model. The motivation for using such a model relates to the fact that devices that are close together are also likely to---in addition to sense correlated signals---harvest correlated amounts of energy. Further, these assumptions are also motivated by the following: \emph{i}) many energy sources, e.g., radio frequency (RF) energy and vibration energy, are known to exhibit exponential decay, which depends on the path-loss in RF signal propagation; therefore, under the assumption of an RF source (or vibration source) and devices located at various distances around it, both $\xi_c^H(i)$ and $\xi_k^H(i)$ would be exponentially decaying; \emph{ii}) the instantaneous operational state of the physical energy converter circuitry of every device is independent from that of other devices~\cite{vullers2010energy,pinuela2012current}; \emph{iii}) the energy source can be modeled as a memoryless process since the energy availability for both RF~\cite{vullers2010energy} and vibration harvesting~\cite{pinuela2012current} fluctuate randomly across time. Overall, our modelling approach is expected to capture key elements of the EH process, in addition to retaining some degree of analytical tractability.

\section{Analysis: Lower Bounds to the Probability of Incorrect Data Reconstruction}
Via lower bounds to the PIDR (i.e., the probability of failure to reconstruct the data captured by all the devices at the BS), we compare the performance of the proposed DCS scheme to that of conventional CS data acquisition schemes. The PIDR associated with the data gathering approaches can be lower bounded by the probability that the energy availability at the devices is not sufficient to fit the energy consumption requirements. These energy consumption requirements are in turn dictated by the set of conditions on the number of measurements at the various devices necessary for successful CS or DCS data reconstruction at the BS (see Appendix A).

\newtheorem{theorem}{Theorem}
\begin{theorem}\label{thm:management1_Proposition}
The PIDR under the proposed signal and EH models for CS and DCS data acquisition can be lower bounded in any activation interval as follows: \footnote{The results of the proposed approach do not depend on the activation index $i$, so, in this section, we drop this index to simplify the notation.}
\begin{equation}\label{eq:management1_Proposition_2}
\begin{split}
&\text{PIDR}_\texttt{CS}
\geq 1-\frac{\sum_{k=1}^{K}\lambda_k e^{-\lambda_c s\tau}}{\sum_{k=1}^{K}\lambda_k-\lambda_c} +
\frac{\lambda_c e^{-\sum_{k=1}^{K}\lambda_k s\tau}}{\sum_{k=1}^{K}\lambda_k-\lambda_c} ,
\end{split}
\end{equation}
and
\begin{equation}\label{eq:management1_Proposition_3}
\begin{split}
&\text{PIDR}_\texttt{DCS} \geq 1\!-\!\min\Bigg\{\!\frac{\sum_{k=1}^{K}\lambda_k e^{-\lambda_c s'\tau}}{\sum_{k=1}^{K}\lambda_k-\lambda_c} -
\frac{\lambda_c e^{-\sum_{k=1}^{K}\lambda_k s'\tau}}{\sum_{k=1}^{K}\lambda_k-\lambda_c} ,\\
&e^{-\lambda_c(\frac{s_c'\tau}{K}+s'\tau)}\! +\! \sum\limits_{k=1}^{K}
\frac{\lambda_c \big(e^{-\lambda_c(\frac{s_c'\tau}{K}+s'\tau)}\! -\! e^{-\lambda_k (s_c'\tau+Ks'\tau)}\big)}{(K\lambda_k-\lambda_c)\prod_{j=1,j\neq k}^{K}(1-\lambda_k/ \lambda_j)}
\Bigg\}.
\end{split}
\end{equation}
\end{theorem}

\begin{IEEEproof}
See Appendix B.
\end{IEEEproof}

\par
The lower bounds to the PIDR embody various attributes associated with the performance of the various data gathering schemes. One can immediately infer from the lower bound in (\ref{eq:management1_Proposition_2}) that the performance of CS based data acquisition tends to deteriorate with the increase in the number of devices $K$, the increase in the signal sparsity $s$, and the decrease in mean energy availability $\frac{1}{\lambda_c}$ or $\frac{1}{\lambda_k}$ ($k=1,\ldots,K$). One can also infer additional behavior associated with the lower bounds by conducting an asymptotic analysis---using Taylor series expansions---in the regime where the EH process is highly correlated across the devices ($\lambda_k \to \infty$) ($k=1,\ldots,K$) and in the regime where the EH process is highly uncorrelated across the devices ($\lambda_c \to \infty$).

\par
When the EH process is highly correlated, i.e., $\lambda_k \to \infty$ ($k=1,\ldots,K$) and $\lambda_c$ is finite, the lower bounds to the PIDR can be expanded as follows:
\begin{equation}\label{eq:management1_cs_1}
\text{PIDR}_\texttt{CS}
\geq 1-e^{-\lambda_c s\tau}+\mathcal{O}\Big(1/\sum_{k=1}^{K}\lambda_k\Big),
\end{equation}
\begin{equation}\label{eq:management1_dcs_1}
\begin{split}
\text{PIDR}_\texttt{DCS} 
\geq1-e^{-\frac{\lambda_c}{K} (s_c'\tau+K s'\tau)}+\mathcal{O}\Big(1/\sum_{k=1}^{K}\lambda_k\Big).
\end{split}
\end{equation}
We can thus conclude via (\ref{eq:management1_cs_1}) and (\ref{eq:management1_dcs_1}) that:
\begin{itemize}
  \item The mean available energy per device, which is given by $1/\lambda_c$, dramatically affects the performance of both data acquisition methods. In particular, the lower bounds to the PIDR in (\ref{eq:management1_cs_1}) and (\ref{eq:management1_dcs_1}) now increase exponentially to unity with the increase in $\lambda_c$.
  \item The signal sparsities also affect the performance of CS and DCS data acquisition considerably. Since $s \approx s'_c + s' \geq s'_c/K + s'$ one concludes that the lower bound in (\ref{eq:management1_cs_1}) is higher than the lower bound in (\ref{eq:management1_dcs_1}).
  \item The network size, as expected, does not affect the lower bounds associated with CS data acquisition (since the signals are reconstructed independently); in contrast, the network size affects the lower bound associated with DCS data acquisition via the common signal component (since the signals are reconstructed simultaneously). In view of the fact that $s \approx s_c'+s' \geq s'_c/K + s'$ one can immediately conclude that the lower bound in (\ref{eq:management1_dcs_1}) can be much higher than the lower bound in (\ref{eq:management1_cs_1}) for a network with a large number of nodes (particularly when $s'_c \gg s'$).

\end{itemize}

\par
In contrast, when the EH process is highly uncorrelated, i.e., $\lambda_c \to \infty$ and $\lambda_k$ ($k = 1,\ldots,K$) are finite, the lower bounds to the PIDR can be expanded as follows:
\begin{equation}\label{eq:management1_cs_2}
\text{PIDR}_\texttt{CS}
\geq 1- e^{-\sum_{k=1}^{K}\lambda_k s\tau} +\mathcal{O}(1/\lambda_c),
\end{equation}
\begin{equation}\label{eq:management1_dcs_2}
\begin{split}
\text{PIDR}_\texttt{DCS} \geq &\max\!\Big\{\!1\!-\!e^{-\sum_{k=1}^{K}\lambda_k s'\tau}, 1\!-\!\sum\limits_{k=1}^{K}
\frac{e^{-\lambda_k (s_c'\tau+Ks'\tau)}}{\prod_{j=1,j\neq k}^{K}(1\!-\frac{\lambda_k}{ \lambda_j})}
\!\Big\}\\
&+\mathcal{O}(1/\lambda_c).
\end{split}
\end{equation}
We can also conclude via (\ref{eq:management1_cs_2}) and (\ref{eq:management1_dcs_2}) that:

\begin{itemize}
  \item The mean available energy per device, which is now given by $1/\lambda_k$ ($k=1,\ldots,K$), also dramatically affects the performance of both data acquisition methods. In particular, the lower bounds to the PIDR in (\ref{eq:management1_cs_2}) and (\ref{eq:management1_dcs_2}) now increase rapidly to unity with the increase in $\lambda_k$ ($k=1,\ldots,K$).
  \item The signal sparsity affects the performance of CS and DCS data acquisition. As $s>s'$, the lower bound associated with CS data acquisition is higher than the first term of the lower bound associated with DCS. In addition, as $K s \approx K(s'_c + s') \geq s'_c + Ks'$, the lower bound associated with CS data acquisition, which results from $1-\text{Pr}\left(\xi_1\geq s\tau,\ldots,\xi_K\geq s\tau\right)$, is also higher than the second term of the lower bound associated with DCS, which results from $1-\text{Pr}\bigg(\sum\limits_{k=1}^{K}\xi_k\geq s_c'\tau+Ks'\tau\bigg)$.
  \item The behavior of the performance of CS and DCS data acquisition as a function of the network size is more interesting in the highly uncorrelated than in the correlated EH scenario. In particular, the lower bound associated with CS data acquisition in (\ref{eq:management1_cs_2}) rapidly tends to unity with increasing network size. In contrast, the behavior of the lower bound associated with DCS data acquisition in (\ref{eq:management1_dcs_2}) depends on the interplay between the two terms in the argument of the $\max(\cdot,\cdot)$ function: the first term tends to increase with the increase in $K$, but the second term, which coincides with the cumulative distribution function (CDF) of a generalized Erlang distributed random variable with mean value $\sum_{k=1}^{K}\frac{1}{\lambda_k(s_c'\tau+K s'\tau)}$, could decrease with the increase in $K$. One then infers that there may be an optimal network size for DCS based data acquisition in the highly uncorrelated EH scenario.
\end{itemize}

\begin{figure}[!t]%
\centering%
\includegraphics[width=0.5\textwidth]{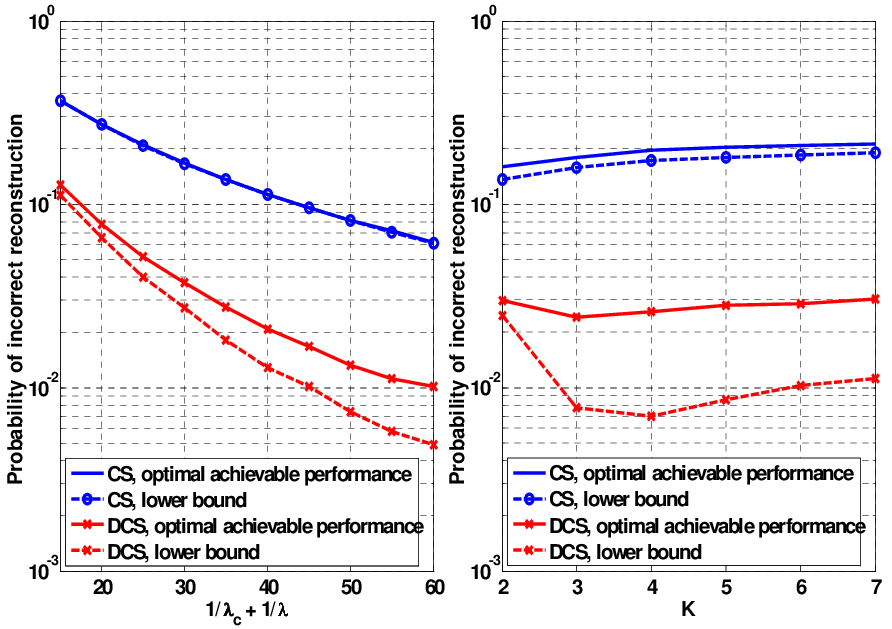}%
\DeclareGraphicsExtensions. \caption{Comparison of the lower bounds with the optimal achievable performance ($1/\lambda = 1/\lambda_1=\ldots=1/\lambda_K$, $\lambda/\lambda_c=5$, $\tau=1$ and $n=50$). The left sub-figure corresponds to $K=2$, $s'=1$ and $s_c'=5$; The right sub-figure corresponds to $s'=1$, $s_c'=7$ and $1/\lambda_c+1/\lambda=40$.} \label{fig:bounds}
\end{figure}%
\par
Finally in Fig.~\ref{fig:bounds}, we give a comparison of the lower bounds with the optimal achievable performance\footnote{The generation of the data is the same as the synthetic experiments given in Section \uppercase\expandafter{\romannumeral4}.}. The optimal achievable performance is obtained directly from the sufficient conditions for successful reconstruction in Appendix A by using Monte Carlo simulations. In addition, numerical results both with synthetic and real data in the sequel reveal that our lower bounds also embody the main performance trends, hence can be used to gauge core issues surrounding the effect of various system phenomena and parameters. In particular, they show the fact that the DCS acquisition and reconstruction approach, in view of its ability to strike a trade-off between the number of measurements taken at different devices without compromising data reconstruction quality, offers the means to match the energy demand to the random nature of the energy supply in order to increase the lifetime and/or the data gathering capability of the network. For example, the left-hand sub-figure shows that CS requires two times more average energy than DCS for networks consisting of two devices to achieve a target PIDR of $10^{-1}$.

\section{Experimental Results}
We now illustrate the potential of the approach both with synthetic data as well as with real dataset~\cite{intellabWSN}. We retain the previous synthetic EH model in both instances. We compare our DCS approach against CS as well as the DSC system~\cite{deligiannis2014maximum}.

\subsection{DCS vs. CS}
In the experiments with synthetic data, we generate sparse signal representations $\mathbf{x}_k$ ($k=1,\ldots,K$) obeying the SCCI model, where the innovation components of various signals exhibit the same support size. Both the common component support and the innovation component supports are selected randomly, and the non-zero elements in the common component and innovation components are drawn independent and identically distributed (i.i.d.) from a Gaussian distribution with zero mean and unit variance. We also generate the equivalent sensing matrices $\mathbf{A}_k$ ($k=1,\ldots,K$) randomly with elements drawn i.i.d. from a zero mean and unit variance Gaussian distribution. The EH process obeys the proposed correlated EH model, where the common component of the harvested energy across the devices follows an exponential distribution with a pre-specified mean $1/\lambda_c$ and the innovation component of the harvested energy per device are drawn from i.i.d. exponential distributions with the same mean $1/\lambda = 1/\lambda_1=\ldots=1/\lambda_K$. We use the CVX package to reconstruct the signals for the CS case in (\ref{eq:CS_L1}) and the DCS case in (\ref{eq:CS_DCS}).


\begin{figure}[!t]%
\centering%
\includegraphics[width=0.5\textwidth]{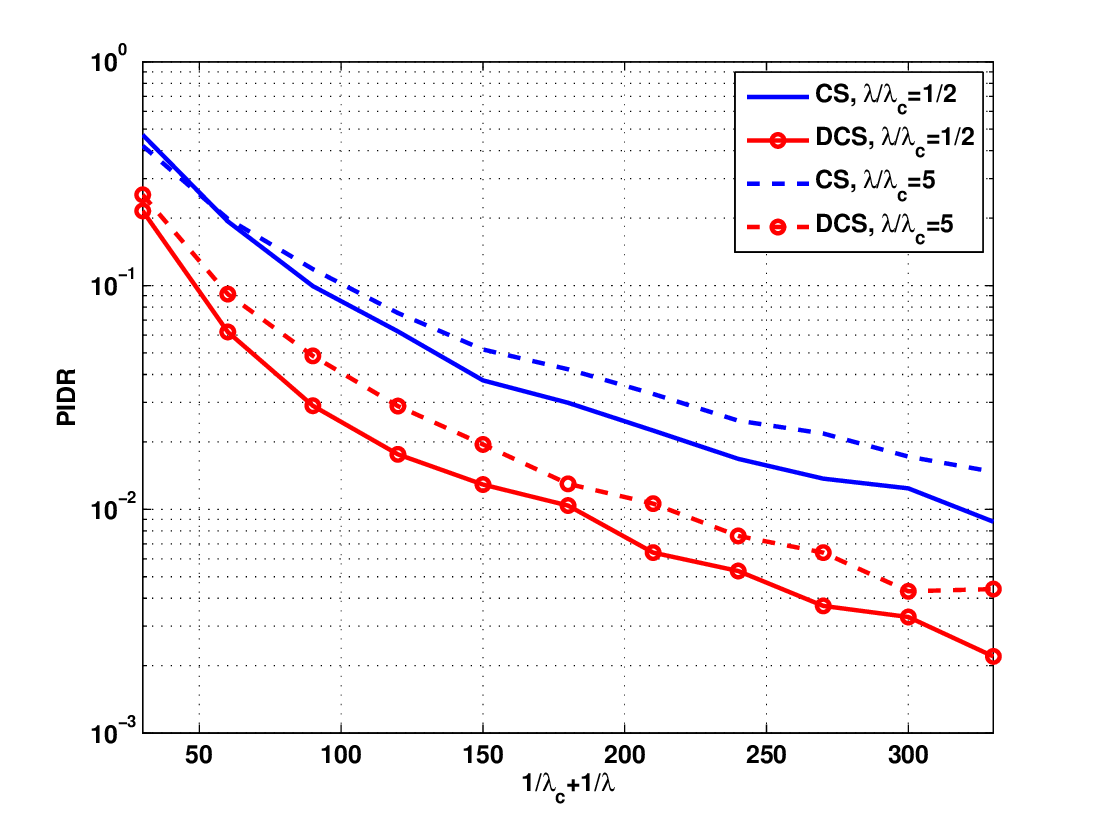}%
\DeclareGraphicsExtensions. \caption{Probability of incorrect reconstruction vs. average harvested energy per device $1/\lambda_{c}+1/\lambda$ ($K=2$, $\tau=1$, $n=50$, $s'=1$ and $s_c'=4$).} \label{fig:prob-lambda-com}
\end{figure}%

\par
Fig.~\ref{fig:prob-lambda-com} shows the PIDR versus average harvested energy per device (i.e., $1/\lambda_c+1/\lambda$) for different ratios between the average energy of the common component and the innovation component (i.e., $\frac{\lambda}{\lambda_{c}}$). As expected, the performance improves with the increase in average harvested energy per device for all schemes. As predicted by our analysis, we observe that DCS performs better than CS for both the less correlated EH scenario ($\lambda/\lambda_c = 5$) and the more correlated EH scenario ($\lambda/\lambda_c = 1/2$). These trends are due to the fact that DCS is able to adapt to the energy variability across the devices whereas CS cannot perform such adaptation. It is also interesting to note that---even though Fig.~\ref{fig:prob-lambda-com} appears to suggest that the performance in the more EH correlated scenario tends to be better than that in the less EH correlated scenario---there appears to be a $\lambda/\lambda_c$ value that leads to the best performance, as shown in Table~\ref{table:prob-ratio-com}.

\begin{table}[t]
\caption{The PIDR for two devices with different ratios between average value of the common energy component and average value of the innovation energy component ($\tau=1$, $s'=1$, $s_c'=4$ and $n=50$).} 
\centering 
\begin{tabular}{c c c c c c} 
\hline\hline 
& & $\frac{\lambda}{\lambda_c}=0$ & $\frac{\lambda}{\lambda_c}=\frac{2}{5}$ & $\frac{\lambda}{\lambda_c}=\frac{4}{3}$ & $\frac{\lambda}{\lambda_c}=\infty$\\[0.5ex]
\hline 
CS& $\frac{1}{\lambda}+\frac{1}{\lambda_c}=200$ & 0.1469 & 0.0261 & 0.0218 & 0.0859 \\
DCS& $\frac{1}{\lambda}+\frac{1}{\lambda_c}=200$ & 0.0599 & 0.0069 & 0.0064 & 0.0740\\
CS& $\frac{1}{\lambda}+\frac{1}{\lambda_c}=300$ & 0.0992 & 0.0129 & 0.0112 & 0.0552\\
DCS& $\frac{1}{\lambda}+\frac{1}{\lambda_c}=300$ & 0.0471 & 0.0035 & 0.0026 & 0.0489\\
\hline 
\end{tabular}
\label{table:prob-ratio-com} 
\end{table}



\begin{figure}[!t]%
\centering%
\includegraphics[width=0.49\textwidth]{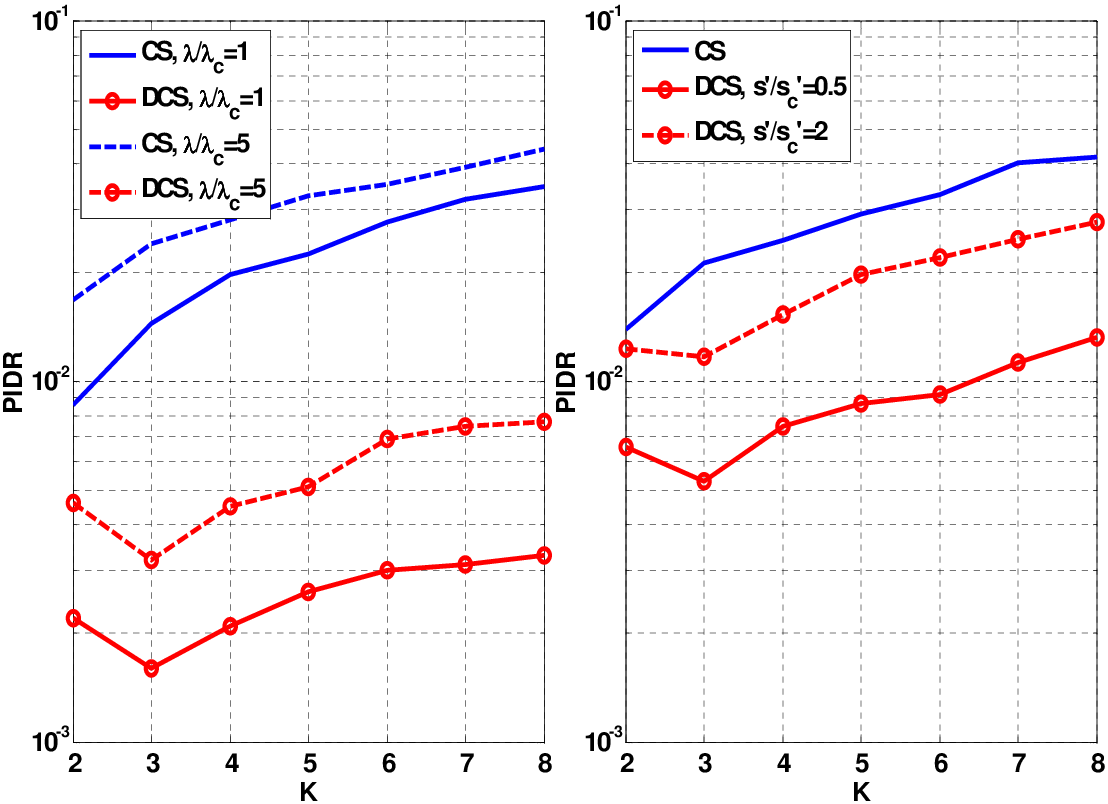}%
\DeclareGraphicsExtensions. \caption{Probability of incorrect data reconstruction vs. number of devices $K$ ($\tau=1$ and $n=50$ ). The left sub-figure corresponds to $s'=1$, $s_c'=4$ and $1/\lambda_c+1/\lambda=300$; the right sub-figure corresponds to $1/\lambda=150$, $1/\lambda_c=150$ and $s'+s_c'=6$.} \label{fig:prob-k}
\end{figure}%


\par
Fig.~\ref{fig:prob-k} shows the PIDR versus the number of devices for different ratios between the average values of the common and the innovation energy components (left hand figure) and for different ratios between the sizes of the signal innovations component support and the signal common component support (right hand figure). We confirm that the PIDR for the DCS approach first decreases and then increases with the number of devices. In contrast, the PIDR for CS increases as the number of devices grows. In addition, the presence of an optimal number of devices for the DCS-based approach is more pronounced in the high signal correlation than in the low signal correlation case.

\par
Table~\ref{table:prob-target} illustrates the average harvested energy per device required to achieve a target PIDR of $10^{-2}$ for different network sizes and different ratios between the average values of the common and the innovation energy components. It is clear that DCS requires much less energy than CS based data gathering and reconstruction. It is observed that the gain of the DCS approach tends to increase with the size of the network. For example, CS requires two times more average energy than DCS for networks consisting of two devices, while for networks consisting of eight devices, CS requires six times more average energy than DCS. Since the amount of harvested energy is a function of the devices' duty cycle, using the proposed DCS approach can increase the duty cycle of devices by approximately six times in comparison to the CS scheme for a network consisting of eight devices, and thus can increase the data gathering rate six times approximately.

\setlength{\tabcolsep}{2pt}
\begin{table}[t]
\caption{The average harvested energy per device required for a target PIDR of $10^{-2}$ ($\tau=1$, $s'=1$, $s_c'=4$ and $n=50$) .} 
\centering 
\begin{tabular}{c c c c c} 
\hline\hline 
& CS & DCS & CS & DCS\\[0.5ex] 
& $\lambda=\lambda_c$ & $\lambda=\lambda_c$ & $\lambda=2\lambda_c$ & $\lambda=2\lambda_c$\\
\hline 
$K=2$ & 330 & 160 & 420 & 215 \\ 
$K=5$ & 560 & 140 & 570 & 155 \\ 
$K=8$ & 1000& 160 & 1100& 180 \\
\hline 
\end{tabular}
\label{table:prob-target} 
\end{table}

We now consider the temperature data collected by the Intel-Berkeley Research Lab~\cite{intellabWSN}---in particular, we consider the contiguous temperature data available from $8$ devices, namely, sensor 1, 2, 3, 4, 7, 8, 9, 10. In order to carry out EH and energy consumption calculations, we assume that each device is equipped with a solar panel with an average harvesting capability of $10\mu \texttt{W/cm}^2$ for the indoor environment given in~\cite{roundy2004power}. We also assume that the harvested power is exponentially distributed with $\frac{1}{\lambda_c}=\frac{1}{\lambda_k}=5\mu \texttt{W/cm}^2$ ($k=1,\ldots,K$). To quantify the energy consumed during transmission, we consider the use of a typical 250$\texttt{kbps}$ 62.64$\texttt{mW}$ ($17.4\texttt{mA} \times 3.6\texttt{V}$) ZigBee RF transceiver. To simplify our comparisons, we ignore the sensing energy cost in this investigation as transmission energy is known to be much higher than the energy cost in compressive non-uniform random sampling~\cite{6155205}. Prior to transmission, each compressive measurement is discretized to 8 bits using a uniform quantizer. Under this setting, the energy required to transmit one measurement is $\tau = \frac{62.64\times 8}{250}\times10^{-6}\frac{J}{\text{bit}}=2.00448\frac{\mu J}{\text{measurement}}$.
The devices independently and randomly collect a small portion of the original samples, quantize them, and then transmit them to the BS based on the available energy. The temperature signals have length $n=397$. Note that the monitored temperature signals are compressible (rather than exactly sparse) in the discrete cosine transform (DCT) domain.

\begin{figure}[!t]%
\centering%
\includegraphics[width=0.5\textwidth]{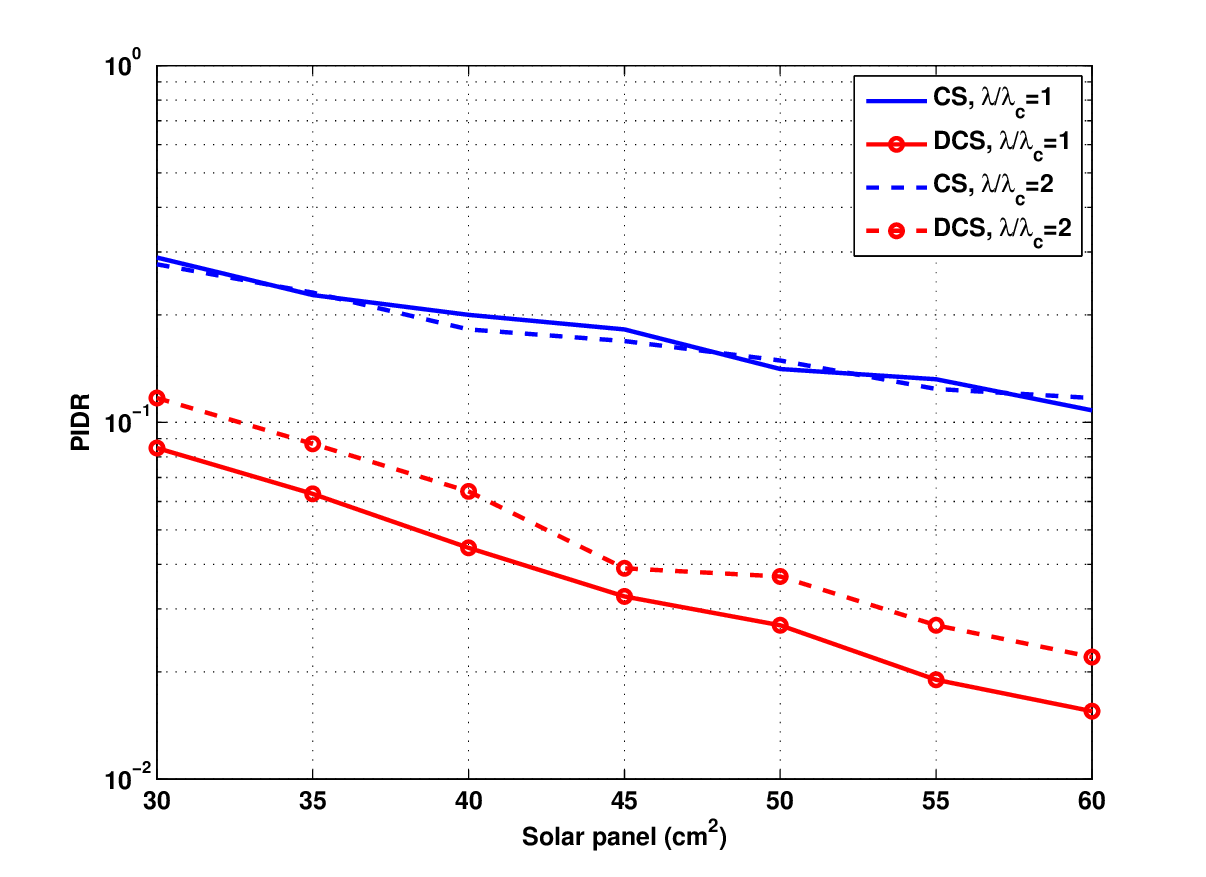}%
\DeclareGraphicsExtensions. \caption{Probability of incorrect data reconstruction vs. solar panel size ($K=2$).} \label{fig:prob-real-lambda}
\end{figure}%

\begin{figure}[!tb]%
\centering%
\includegraphics[width=0.5\textwidth]{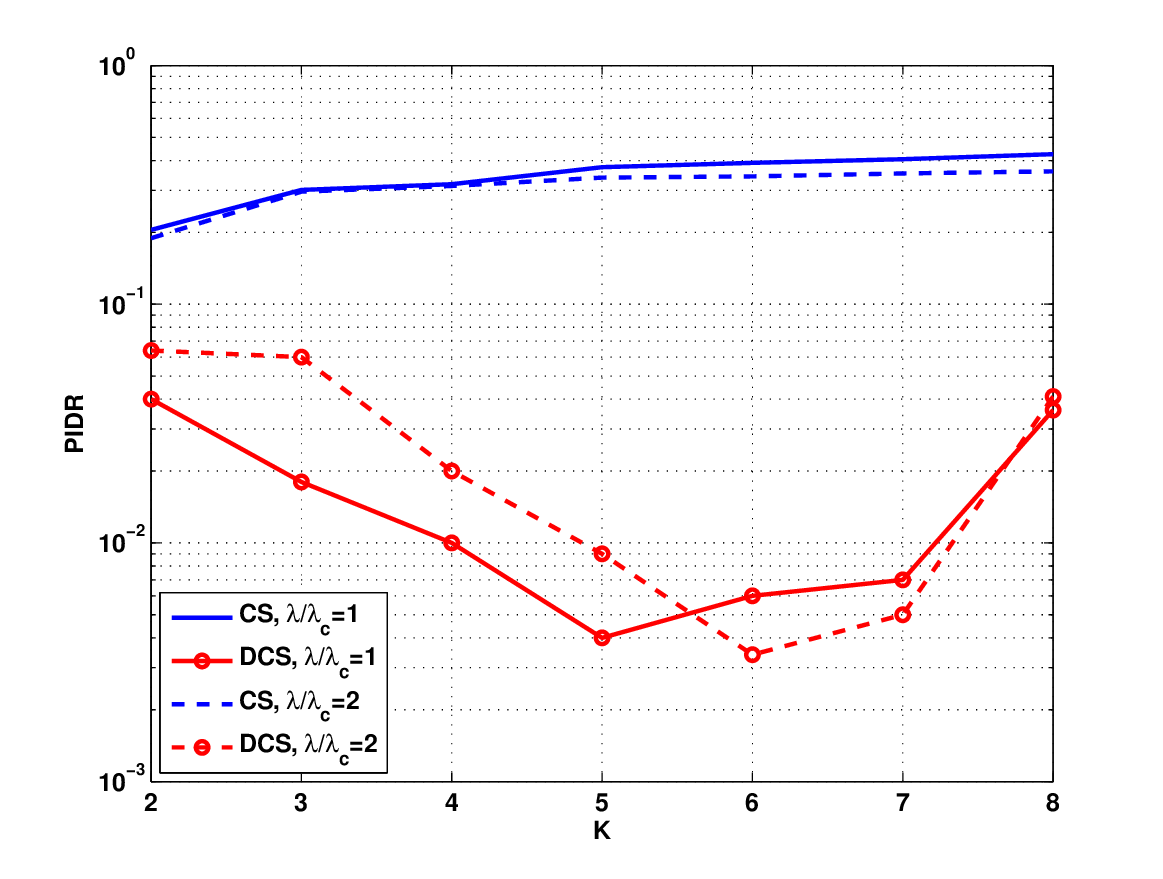}%
\DeclareGraphicsExtensions. \caption{Probability of incorrect reconstruction vs. number of devices $K$ (with a 40$\texttt{cm}^2$ solar panel).}
\label{fig:prob-real-k}
\end{figure}%

We compare the proposed DCS-based approach versus the baseline CS-based system. We assume that the reconstruction is successful if the relative recovery error for a single device satisfies $\frac{\|\hat{\mathbf{f}}_k-\mathbf{f}_k\|_{2}^2}{\|\mathbf{f}_k\|_{2}^2}< 10^{-3}$,
where $\mathbf{f}_k$ and $\hat{\mathbf{f}}_k$ denote the original signal and the reconstructed signal of the $k$th device, respectively.
Fig.~\ref{fig:prob-real-lambda} shows the PIDR for $K=2$ devices (i.e., sensor 2 and 3), achieved by the CS and the proposed DCS data gathering schemes for various solar panel sizes. It is clear that the DCS scheme requires much lower energy levels in comparison to CS for a certain target PIDR. For example, with $\lambda/\lambda_c=1$, achieving a PIDR equal to $10^{-1}$, requires the devices to be equipped with solar panels of size 30$\texttt{cm}^2$ and 60$\texttt{cm}^2$ when using the proposed DCS and the conventional CS-based approach, respectively. It is evident that using the proposed approach can considerably ease the EH capability requirements per device.

\par
Considering various numbers of devices communicating correlated data to a BS, Fig.~\ref{fig:prob-real-k} shows the PIDR with a solar panel of fixed size achieved by the CS and the proposed DCS schemes. In contrast to the conventional CS-based approach, the DCS-based scheme achieves a lower PIDR. In addition, the PIDR for the DCS approach first decreases and then increases with the number of the devices. This result highlights the capacity of DCS to exploit both intra- and inter-sensor correlations in the gathered data. There could be many factors that determine the best $K$ achieving the lowest PIDR. For example, the real world signal is not exactly sparse but rather nearly sparse; there are approximation errors in the SCCI model for charactering the inter-sensor correlations. We remark that the settings behind \Cref{fig:prob-lambda-com,fig:prob-k,fig:prob-real-lambda,fig:prob-real-k} are such that the device is powered only via the energy harvested from the environment. As such, the fact that the DCS-based approach exhibits higher energy efficiencies in data collection forms the basis of our energy neutrality claims.

\begin{figure}[!tb]%
\centering%
\includegraphics[width=0.45\textwidth]{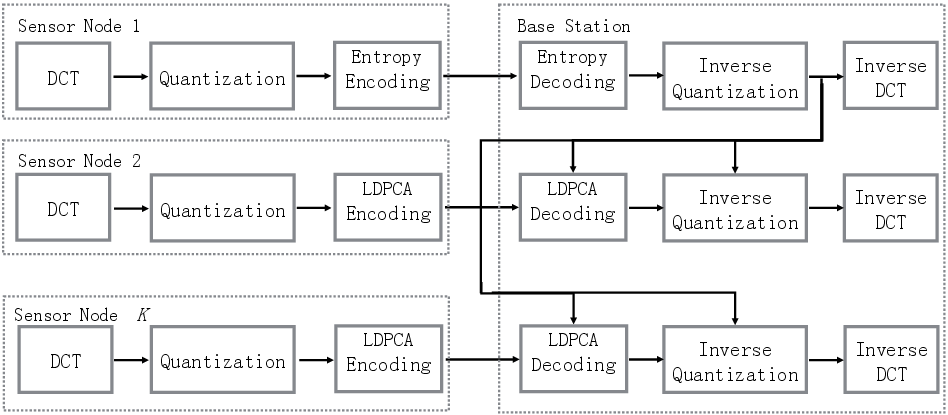}%
\DeclareGraphicsExtensions. \caption{The considered coding architecture that performs distributed source coding by means of Wyner-Ziv coding.}
\label{fig:WZcodingArchitecture}
\end{figure}%

\subsection{DCS vs. Distributed Source Coding (DSC)}
We now compare the proposed DCS scheme against a DSC system \cite{deligiannis2014maximum} that performs efficient compression of the correlated data collected by the devices, as shown in Fig. \ref{fig:WZcodingArchitecture}. The experimental datasets~\cite{intellabWSN}, as well as the energy harvesting and consumption profiles are as in the previous section\footnote{The assumption that the encoding complexity for the DSC and DCS are comparable is made based on the following observations: i) DCT is an extra operation with respect to the DCS encoder; ii) quantization is the same as in DCS (the only difference w.r.t. DCS is that it is applied before dimentionality reduction, therefore more samples are quantised than in DCS); LDPCA encoding has a comparable complexity with multiplication with a random matrix (performed in DCS).}. The benchmark DSC system is based on the principles of Wyner-Ziv coding \cite{wynerziv,deligiannis2014maximum}: namely, the data collected from one device is intra encoded and communicated to the BS (decoder) where it forms the side information used to decode the data from the other devices. It is worth mentioning that, when 2 devices are connected to a BS, Wyner-Ziv coding is optimal in terms of DSC performance. When, however, more than 2 devices are connected to a BS then a DSC scheme based on Berger-Tung coding, i.e., multiterminal source coding, is more efficient. However, multiterminal source coding is not fully characterized in terms of performance bounds and is difficult to implement in practice, especially when the number of devices connected to a BS increases.


\begin{figure*}
\centering
\subfigure[]{
\includegraphics[scale=0.45,trim=0.5cm 0.2cm 0.5cm 0.5cm]{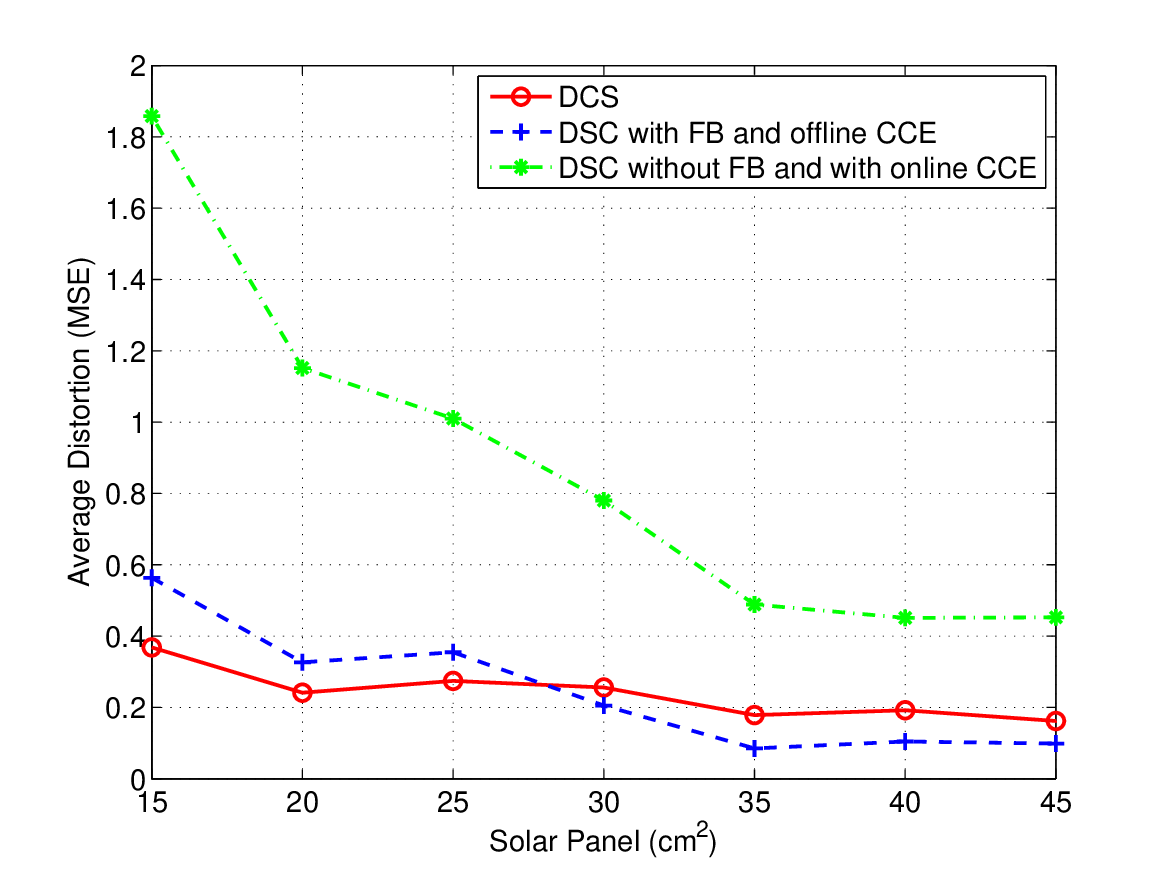}}
\subfigure[]{
\includegraphics[scale=0.45,trim=0.5cm 0.2cm 0.5cm 0.5cm]{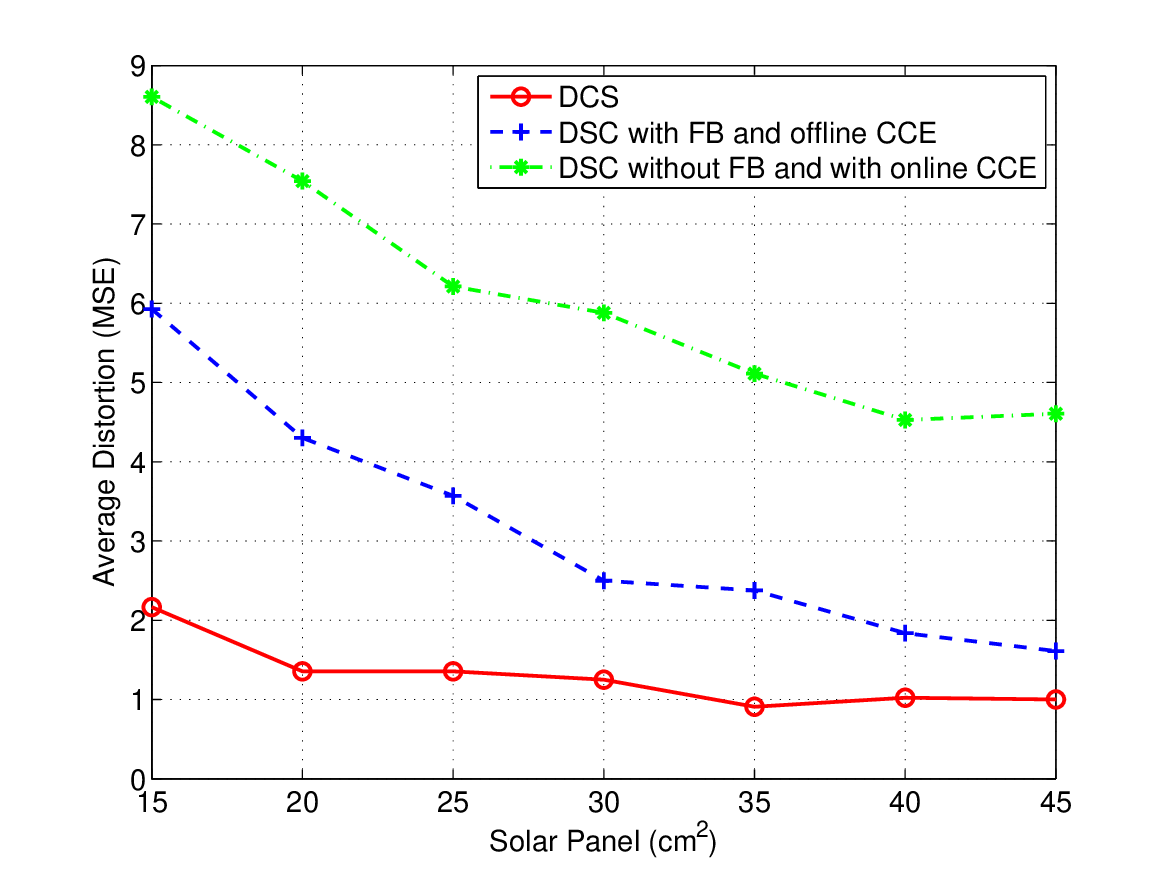}}
\caption{Average MSE distortion versus solar panel size obtained with the
proposed DCS scheme and different DSC configurations, when four devices are connected
to a BS. Temperature data from (a) devices: 1, 2, 3, and 4; (b) devices: 7, 8, 9, and
10, in the Intel-Berkeley database \cite{intellabWSN}.
  \label{fig:4SNsDSCvsDCSResults}}
\end{figure*}

\par
According to the devised DSC architecture \cite{deligiannis2014maximum}, $n=397$ samples are collected by each device and aggregated for encoding. The samples first undergo a DCT to perform intra-sensor data decorrelation. The value of the first DCT coefficient, i.e., the DC coefficient, from each device is binarized and transmitted. The remaining $396$ AC coefficients undergo uniform quantization and the resulting quantization indices are split into bit-planes. At the device performing intra-signal encoding the bit-planes are arithmetic entropy encoded sequentially starting from the most significant one. At the devices performing Wyner-Ziv coding, the bit-planes are Slepian-Wolf \cite{slepianwolf} encoded using the state-of-the-art Low-Density Parity-Check Accumulate (LDPCA) codes \cite{varodayan2006rate}. Concerning Wyner-Ziv rate control, we consider: \textit{(i)} a decoder-driven mechanism deploying a feedback channel to request extra information from the encoder when decoding fails, or \textit{(ii)} an encoder-based scheme as in \cite{verbist2013encoder}. The former performs optimal rate control but suffers from structural delays, while the latter follows a more realistic approach but occasionally fails to accurately estimate the required rate for decoding, thus leading to loss in performance.
\par
For each device---performing either intra-signal coding or Wyner-Ziv coding---the number of encoded and transmitted bit-planes depends on the required encoding rate and the available harvested energy. When the available portion of the harvested energy is not depleted during transmission (because the encoding rate is lower than the available transmission rate), the residual energy is stored in the battery and used during the subsequent data transmission.
\par
At the BS, which runs the decoder, the entropy-encoded bit-planes are first decoded and then compiled into quantization indices. Recall that this information corresponds to the quantized AC coefficients of the data from the device performing intra-signal encoding. After inverse quantization, we use this data as side information to decode the LDPCA-encoded bit-planes of the AC coefficients from the remaining devices. The soft-information required for LDPCA decoding is derived by assuming a correlation channel, where the noise follows a zero-mean Laplace distribution \cite{deligiannis2014maximum}. The scaling parameter of the correlation noise distribution can be derived \textit{(i)} in an offline manner, or \textit{(ii)} using an online technique as in \cite{deligiannis2014maximum}. When the bit-planes are decoded, inverse quantization is performed to obtain the decoded AC coefficient. The AC coefficients from the data-block of each device are then combined with their DC coefficient and inverse DCT is performed to derive the decoded temperature data.

We compare this Wyner-Ziv coding system \cite{deligiannis2014maximum} against the proposed DCS scheme in terms of the mean-squared error (MSE) distortion of the decoded data versus the available harvested energy---expressed through the panel size. We abide by the previous EH model, where the harvested power is exponentially distributed with
$\frac{1}{\lambda_c}=\frac{1}{\lambda_k}=5\mu \texttt{W/cm}^2$, $\forall k=1,\ldots,K$. We report average results over $10^{3}$ independent runs.

Fig. \ref{fig:4SNsDSCvsDCSResults} depicts the results when four devices are connected to a BS. Fig. \ref{fig:4SNsDSCvsDCSResults}(a) shows that the proposed DCS scheme significantly outperforms the practical Wyner-Ziv coding configuration and achieves a performance similar to the optimal DSC system, where FB and CCE denote feedback channel and correlation channel estimation, respectively. However, when the data from the device performing intra-signal encoding (which form the side information) is not highly correlated with the data from all the other devices (which apply Wyner-Ziv coding), then the proposed DCS system significantly outperforms even the optimal (yet impractical) DSC system [see Fig. \ref{fig:4SNsDSCvsDCSResults}(b)]. In particular, as shown in Fig. \ref{fig:4SNsDSCvsDCSResults}(b), the reported reduction in the MSE reduction with respect to the optimal DSC system can mount up to 66.67\% at low EH levels. These performance improvements highlight the capability of the proposed approach to effectively exploit both intra- and inter-sensor data correlations with respect to the state-of-the-art DSC solution \cite{deligiannis2014maximum,1328091}.

\section{Conclusion}
We have proposed a novel DCS-based data acquisition and reconstruction scheme that offers the means to match the energy demand to the energy supply for EH IoT. We have shown that our solution delivers substantial gains in energy efficiency for a certain target data reconstruction quality in comparison to \textit{(i)} a CS-based data acquisition and reconstruction approach, and \textit{(ii)} a DSC system that realizes practical Wyner-Ziv coding. Significant data-reconstruction-versus-energy gains are achieved that translate immediately into improvements in network lifetime and network data gathering capability.

\par
The potential of the proposed DCS-based data acquisition and reconstruction solution to unlock energy neutrality has been unveiled in a setting involving a centralized EH IoT architecture and two basic models: (1) a signal model that captures the fact that the signals collected by different devices exhibit correlation; and (2) a EH model that also captures the fact that the energy harvested by different devices also exhibits some degree of correlation. One would expect some of the key trends to generalize to other correlated signal models and correlated EH models.



%
\appendices

\section{The Necessary and Sufficient Conditions for DCS Reconstruction}
\par
The basis of our analysis are necessary conditions for the successful reconstruction of compressively sensed signals that obey the SCCI model. These necessary conditions along with sufficient conditions, which have been put forth in~\cite{6502243}, are reviewed here.

\par
Let us write
\begin{equation}\label{eq:SCCI_model2}
\tilde{\mathbf{x}}(i)=\mathbf{P}(i)\boldsymbol{\theta}(i),
\end{equation}
where $\tilde{\mathbf{x}}(i)=\left[\mathbf{x}_1(i)^T\ \ldots\
\mathbf{x}_K(i)^T\right]^T\in \mathbb{R}^{Kn(i)}$ is the extended sparse signal representation vector, $\boldsymbol{\theta}(i)=[\boldsymbol{\theta}_c(i)^T\ \boldsymbol{\theta}_1(i)^T\ \ldots\  \boldsymbol{\theta}_K(i)^T]^T\in \mathbb{R}^{s_c'(i)+Ks'(i)}$ is a vector with no zero values, $\boldsymbol{\theta}_c(i)\in \mathbb{R}^{s_c'(i)}$, $\boldsymbol{\theta}_k(i)\in \mathbb{R}^{s'(i)}$ ($k=1,\ldots,K$), and $\mathbf{P}(i)\in \mathbb{R}^{Kn(i)\times \left(s_c'(i)+Ks'(i)\right)}$ denotes a location that admits the form:
\[\mathbf{P}(i)=
\begin{bmatrix}
\mathbf{P}_c(i)&\mathbf{P}_1(i)&\mathbf{0}&\mathbf{0}  & \cdots & \mathbf{0}      \\
\mathbf{P}_c(i)&\mathbf{0}&\mathbf{P}_2(i)&\mathbf{0}  & \cdots & \mathbf{0}      \\
\vdots &  &    &                    & \ddots     & \vdots \\
\mathbf{P}_c(i)&\mathbf{0}&\mathbf{0}&\mathbf{0}    & \cdots & \mathbf{P}_K(i)
\end{bmatrix},\]
where $\mathbf{P}_c(i)\in \mathbb{R}^{n(i)\times s_c'(i)}$ and $\mathbf{P}_k(i)\in \mathbb{R}^{n(i)\times s'(i)}$ ($k=1,\ldots,K$) are different submatrices of $n(i)\times n(i)$ identity matrices.


\par
There can be overlap between the support of the common component associated with $\mathbf{P}_c(i)$ and the supports of innovation components associated with $\mathbf{P}_k(i)$ ($k=1,\ldots,K$). The overlap size is defined as follows:

\newtheorem{definition}{Definition}
\begin{definition}[{\cite[Definition 4]{6502243}}]
Let $\mathcal{J}\subseteq \{1,\ldots,K\}$. Consider also $K$ correlated signals with sparse representations $\mathbf{x}_k(i)$ ($k=1,\ldots,K$) that follow the SCCI model with a given location matrix $\mathbf{P}(i)$. The overlap size between the common component support and innovation component supports for all signals $k\in\{\mathcal{J}^C\}$ is defined as
\begin{equation}\label{eq:overlap_size}\nonumber
\begin{split}
&q(\mathcal{J},\mathbf{P}(i)):=\Big|\{j\in\{1,\ldots,n(i)\}:\text{row}\ j\ \text{of}\ \mathbf{P}_c(i)\ \text{has}\\
&\quad \text{nonzero components and } \forall k\notin \mathcal{J},\ \text{row}\ j\ \text{of }  \mathbf{P}_k(i)\ \text{has}\\
&\quad \text{nonzero components}\}  \Big|.
\end{split}
\end{equation}
\end{definition}
Note that $q\big(\{1,\ldots,K\},\mathbf{P}(i)\big)=s_c'(i)$.

\par
The following Theorem gives the sufficient condition for the joint successful reconstruction of the $K$ correlated sparse signals with an algorithm based on an enumerative search over all possible sparse patterns~\cite{6502243}.
\begin{theorem}[{\cite[Theorem 3]{6502243}}]\label{thm:DCS_sufficient_condition}
Let the equivalent sensing matrices $\mathbf{A}_k(i)\in \mathbb{R}^{m_k(i)\times n(i)}$ ($k=1,\ldots,K$) be populated with i.i.d. Gaussian entries. Let also the $K$ correlated signals with sparse representations $\mathbf{x}_k(i)$ ($k=1,\ldots,K$) follow the SCCI model with a full-rank location matrix $\mathbf{P}(i)$. If
\begin{equation}\label{eq:DCS_sufficient_condition}
\begin{split}
\sum\limits_{k\in \mathcal{J}} m_k(i)\geq |\mathcal{J}| s'(i) + q(\mathcal{J},\mathbf{P}(i))+|\mathcal{J}|
\end{split}
\end{equation}
for all subsets $\mathcal{J}\subseteq \{1,\ldots,K\}$, then the $K$ correlated signals can be successfully recovered.
\end{theorem}

\par
The following Theorem now gives a necessary condition for the joint successful reconstruction of the $K$ correlated sparse signals~\cite{6502243}.

\begin{theorem}[{\cite[Theorem 2]{6502243}}]\label{thm:DCS_necessary_condition}
Let the equivalent sensing matrices $\mathbf{A}_k(i)\in \mathbb{R}^{m_k(i)\times n(i)}$ ($k=1,\ldots,K$). Let also the $K$ correlated signals with sparse representations $\mathbf{x}_k(i)$ ($k=1,\ldots,K$) follow the SCCI model with a full-rank location matrix $\mathbf{P}(i)$. If
\begin{equation}\label{eq:DCS_necessary_condition}
\begin{split}
\sum\limits_{k\in \mathcal{J}} m_k(i)< |\mathcal{J}| s'(i) + q(\mathcal{J},\mathbf{P}(i))
\end{split}
\end{equation}
for any subset $\mathcal{J}\subseteq \{1,\ldots,K\}$, there exists a different set of correlated signals with sparse representations that also follow the SCCI model with signal measurements that are identical to those of the original desired signal.
\end{theorem}

%
%

\par
The necessary conditions for successful DCS reconstruction in Theorem~\ref{thm:DCS_necessary_condition} also specialize to the conventional necessary condition for successful CS reconstruction that entails that the number of projections ought to be greater than or equal to the signal sparsity, by taking $K = 1$ and removing the common component.

\par
In addition to the sufficient and necessary conditions for DCS reconstruction embodied in Theorems~\ref{thm:DCS_sufficient_condition} and~\ref{thm:DCS_necessary_condition} respectively, it has been observed that the number of measurements of various signals for DCS can be substantially lower than the number of measurements in CS. In addition, it has also been observed, as hinted at in Theorems~\ref{thm:DCS_sufficient_condition} and~\ref{thm:DCS_necessary_condition}, that the number of measurements of various signals for DCS can also be adjusted without compromising data recovery in practice owing to the inter-signal correlation~\cite{6502243,baron2006distributed}.

\section{Proof of Theorem~\ref{thm:management1_Proposition}}
The analysis requires the distribution of a sum of independent exponential random variables with distinct parameters in some calculations~\cite{akkouchi2008convolution}.
\newtheorem{lemma}{Lemma}
\begin{lemma}
Let $\beta_1,\ldots,\beta_K$ be $K$ independent exponential random variables with distinct parameters $\lambda_1,\ldots,\lambda_K$ respectively. Then, the probability density function of the random variable $\beta = \sum\limits_{k=1}^{K}\beta_k$ is given by:
\begin{equation}\label{eq:lemma_1}
\emph{P}_\beta(t) =
\sum\limits_{k=1}^{K} \frac{\prod_{j=1}^{K}\lambda_j}{\prod_{j=1,j\neq k}^{K}(\lambda_j-\lambda_k)}e^{-\lambda_k t}H(t),\\
\end{equation}
where $H(t)=1$ if $t\geq 0$ and $H(t)=0$ otherwise.
\end{lemma}

\subsection{CS Based Data Acquisition and Reconstruction}
By using the assumptions about the EH process in Section \uppercase\expandafter{\romannumeral2}, the probability of incorrect data collection due to energy depletion for a CS data acquisition scheme can be lower bounded as follows:
\begin{equation}\label{eq:prob_raw_CS}
\begin{split}
&\text{PIDR}_\texttt{CS}
\geq1-\text{Pr}\left(\xi_1\geq s\tau ,\ldots,\xi_K\geq s\tau \right)\\
&=\!1-\!\text{Pr}\left(\hat{\xi}^H_1 \geq s\tau -\hat{\xi}^H_c,\ldots,\hat{\xi}^H_K \geq s\tau -\hat{\xi}^H_c \right)\\
&=\!1\!\!-\!\!\!\int_{s\tau}^{\infty}\!\!\! \emph{P}_{\hat{\xi}^H_c}\!(\!t\!) dt \! -\!\!\! \int_{0}^{s\tau}\!\! \!\!\!\int_{s\tau\! -\!t}^{\infty}\!\!\!\!\!
\cdots \!\!\int_{s\tau\! -\!t}^{\infty}\!\!\! \emph{P}_{\hat{\xi}^H_c}\!(\!t\!)\!\prod\limits_{k=1}^{K}\!\emph{P}_{\hat{\xi}^H_k}\!(\!t_k\!)\!
dt_1\!\cdots\! dt_K dt\\
&=\!1\!-\!\frac{\sum_{k=1}^{K}\lambda_k}{\sum_{k=1}^{K}\lambda_k-\lambda_c} e^{-\lambda_c s\tau}+
\frac{\lambda_c}{\sum_{k=1}^{K}\lambda_k-\lambda_c} e^{-\sum_{k=1}^{K}\lambda_k s\tau}.
\end{split}
\end{equation}
Note that we use the specialization of the necessary condition in Theorem~\ref{thm:DCS_necessary_condition} from the DCS to the CS case that states that the number of measurements per device has to be greater than or equal to the signal sparsity for successful reconstruction.

\subsection{DCS Based Data Acquisition and Reconstruction}
By using the assumptions associated with the EH process in Section \uppercase\expandafter{\romannumeral2} together with the necessary conditions for successful reconstruction in Appendix A, the probability of incorrect data collection due to energy depletion for a DCS data acquisition scheme can be lower bounded as follows:
\begin{subequations}\label{eq:prob_DCS}
\begin{align}
&\text{PIDR}_\texttt{DCS}\! \geq\!1\!-\!\text{Pr}\bigg(\! \xi_1\!\geq\!  \big(\!s'\!+\!q(\!\{1\},\mathbf{P}\!)\big)\!\tau,\ldots, \xi_K\!\geq\!  \big(\!s'\!+\!q(\!\{K\},\mathbf{P}\!)\big)\!\tau,\notag \\
&\qquad  \xi_1+\xi_2\geq \big(2s'+q(\{1,2\},\mathbf{P})\big)\tau,\ldots, \notag\\
&\qquad \sum\limits_{k=1}^{K}\xi_k\geq \Big(Ks'+q(\{1,\ldots,K\},\mathbf{P})\Big)\tau\bigg)\notag\\
&\! \geq\!1\!-\!\text{Pr}\bigg(\!\xi_1\!\geq\!  \big(s'\!+\!q(\{1\},\mathbf{P})\big)\tau,\ldots, \xi_K\!\geq\!  \big(s'\!+\!q(\{K\},\mathbf{P})\big)\tau,\notag\\
&\qquad \sum\limits_{k=1}^{K}\xi_k\geq \big(Ks'+q(\{1,\ldots,K\},\mathbf{P})\big)\tau\bigg) \label{eq:prob_DCS_a}\\
&\! \geq\!1\!-\!\min\!\bigg\{\! \text{Pr}\Big(\!\xi_1\!\geq\!  s'\tau,\ldots,\xi_K\!\geq\!  s'\tau\!\Big), \text{Pr}\Big(\!\sum\limits_{k=1}^{K}\!\xi_k\!\geq \! s_c'\tau\!+\!Ks'\tau\!\Big)\!\bigg\}.\label{eq:prob_DCS_b}
\end{align}
\end{subequations}
where in (\ref{eq:prob_DCS_a}) we loosen the bound in order to reduce the number of conditions on the harvested energy (where equality holds for $K=2$), and in (\ref{eq:prob_DCS_b}) we loosen the bound further in order to drop the dependency on the location matrix. As
\begin{equation}\label{eq:prob_DCS1}
\begin{split}
&\text{Pr}\Big(\!\xi_1\!\geq\!  s'\tau,\ldots,\xi_K\geq  s'\tau\!\Big)\!=\!\frac{\sum\limits_{k=1}^{K}\lambda_k e^{-\lambda_c s'\tau}}{\sum\limits_{k=1}^{K}\lambda_k-\lambda_c}\! -\!
\frac{\lambda_c e^{-\!\sum\limits_{k=1}^{K}\lambda_k s'\tau}}{\sum\limits_{k=1}^{K}\lambda_k-\lambda_c} ,
\end{split}
\end{equation}
and
\begin{subequations}\label{eq:prob_DCS2}
\begin{align}
&\text{Pr}\!\bigg(\!\sum\limits_{k=1}^{K}\!\xi_k\!\geq\! s_c'\tau\!+\!Ks'\tau\!\bigg)\!=\!\text{Pr}\!\bigg(\!K\hat{\xi}_c^H\!+\! \sum\limits_{k=1}^{K}\!\hat{\xi}_k^H\!\geq\! s_c'\tau\!+\!Ks'\tau\!\bigg) \notag\\
&=\int_{\frac{s_c'\tau}{K}+s'\tau}^{\infty} \emph{P}_{\hat{\xi}^H_c}(t) dt+\notag\\
&\qquad \int_{0}^{\frac{s_c'\tau}{K}+s'\tau} \emph{P}_{\hat{\xi}^H_c}(t)\text{Pr}\bigg(\sum\limits_{k=1}^{K}\hat{\xi}_k^H\geq s_c'\tau+Ks'\tau-K\hat{\xi}_c^H\bigg)
 dt \notag\\
&\!=\!e^{\!-\!\lambda_c\!\big(\!\frac{s_c'\tau}{K}\!+\!s'\tau\!\big)}\!\!+\!\!\!\int_{0}^{\frac{s_c'\tau}{K}+s'\tau}\!\!\!\! \lambda_c e^{-\!\lambda_c t}\!
\sum\limits_{k=1}^{K}\! \frac{e^{-\!\lambda_k\! (\!s_c'\tau\!+\!Ks'\tau\!-\!K\hat{\xi}_c^H\!)}\!\!\prod\limits_{j=1}^{K}\!\lambda_j}{\lambda_k\prod_{j=1,j\neq k}^{K}(\lambda_j-\lambda_k)} dt \label{eq:prob_DCS2_c}\\
&\!=\!e^{-\!\lambda_c\!\big(\!\frac{s_c'\tau}{K}+s'\tau\!\big)}\! +\!\! \sum\limits_{k=1}^{K}\!
\frac{\lambda_c\!\Big(\!e^{-\lambda_c(\frac{s_c'\tau}{K}+s'\tau)}\! -\! e^{-\lambda_k (s_c'\tau+Ks'\tau)}\Big)}{(K\lambda_k-\lambda_c)\prod_{j=1,j\neq k}^{K}(1-\lambda_k/ \lambda_j)},\notag
\end{align}
\end{subequations}
where Lemma 1 is used in deriving (\ref{eq:prob_DCS2_c}), then we have
\begin{equation}\label{eq:prob_DCS3}
\begin{split}
&\text{PIDR}_\texttt{DCS} \geq 1\!-\!\min\Bigg\{\!\frac{\sum_{k=1}^{K}\lambda_k e^{-\lambda_c s'\tau}}{\sum_{k=1}^{K}\lambda_k-\lambda_c} -
\frac{\lambda_c e^{-\sum_{k=1}^{K}\lambda_k s'\tau}}{\sum_{k=1}^{K}\lambda_k-\lambda_c} ,\\
&e^{-\lambda_c(\frac{s_c'\tau}{K}+s'\tau)}\! +\! \sum\limits_{k=1}^{K}
\frac{\lambda_c \big(e^{-\lambda_c(\frac{s_c'\tau}{K}+s'\tau)}\! -\! e^{-\lambda_k (s_c'\tau+Ks'\tau)}\big)}{(K\lambda_k-\lambda_c)\prod_{j=1,j\neq k}^{K}(1-\lambda_k/ \lambda_j)}
\Bigg\}.
\end{split}
\end{equation}

\ifCLASSOPTIONcaptionsoff
  \newpage
\fi


\bibliographystyle{IEEEtran}
\bibliography{IEEEabrv,bib_paper}

\begin{thebibliography}{10}
\providecommand{\url}[1]{#1}
\csname url@samestyle\endcsname
\providecommand{\newblock}{\relax}
\providecommand{\bibinfo}[2]{#2}
\providecommand{\BIBentrySTDinterwordspacing}{\spaceskip=0pt\relax}
\providecommand{\BIBentryALTinterwordstretchfactor}{4}
\providecommand{\BIBentryALTinterwordspacing}{\spaceskip=\fontdimen2\font plus
\BIBentryALTinterwordstretchfactor\fontdimen3\font minus
  \fontdimen4\font\relax}
\providecommand{\BIBforeignlanguage}[2]{{%
\expandafter\ifx\csname l@#1\endcsname\relax
\typeout{** WARNING: IEEEtran.bst: No hyphenation pattern has been}%
\typeout{** loaded for the language `#1'. Using the pattern for}%
\typeout{** the default language instead.}%
\else
\language=\csname l@#1\endcsname
\fi
#2}}
\providecommand{\BIBdecl}{\relax}
\BIBdecl

\bibitem{8820755}
L.~{Zhang}, Y.~{Liang}, and D.~{Niyato}, ``{6G} visions: Mobile
  ultra-broadband, super internet-of-things, and artificial intelligence,''
  \emph{China Communications}, vol.~16, no.~8, pp. 1--14, 2019.

\bibitem{8664000}
X.~{Liu} and N.~{Ansari}, ``Toward green {IoT}: Energy solutions and key
  challenges,'' \emph{IEEE Communications Magazine}, vol.~57, no.~3, pp.
  104--110, 2019.

\bibitem{8938189}
X.~{Wang}, Z.~{Ning}, X.~{Hu}, L.~{Wang}, L.~{Guo}, B.~{Hu}, and X.~{Wu},
  ``Future communications and energy management in the internet of vehicles:
  Toward intelligent energy-harvesting,'' \emph{IEEE Wireless Communications},
  vol.~26, no.~6, pp. 87--93, 2019.

\bibitem{8913774}
C.~{Guo}, J.~{Xin}, L.~{Zhao}, and X.~{Chu}, ``Performance analysis of
  cooperative {NOMA} with energy harvesting in multi-cell networks,''
  \emph{China Communications}, vol.~16, no.~11, pp. 120--129,, 2019.

\bibitem{7001194}
W.~D. {Leon-Salas}, ``Low-complexity compression for sensory systems,''
  \emph{IEEE Transactions on Circuits and Systems II: Express Briefs}, vol.~62,
  no.~4, pp. 322--326, 2015.

\bibitem{8777159}
L.~{Zhao}, T.~{Lin}, D.~{Zhang}, K.~{Zhou}, and S.~{Wang}, ``An ultra-low
  complexity and high efficiency approach for lossless alpha channel coding,''
  \emph{IEEE Transactions on Multimedia}, vol.~22, no.~3, pp. 786--794, 2020.

\bibitem{slepianwolf}
D.~Slepian and J.~K. Wolf, ``Noiseless coding of correlated information
  sources,'' \emph{IEEE Transactions on Information Theory}, vol.~19, no.~4,
  pp. 471--480, 1973.

\bibitem{wynerziv}
A.~D. Wyner and J.~Ziv, ``The rate-distortion function for source coding with
  side information at the decoder,'' \emph{IEEE Transactions on Information
  Theory}, vol.~22, no.~1, pp. 1--10, 1976.

\bibitem{7083694}
N.~{Deligiannis}, E.~{Zimos}, D.~M. {Ofrim}, Y.~{Andreopoulos}, and
  A.~{Munteanu}, ``Distributed joint source-channel coding with
  copula-function-based correlation modeling for wireless sensors measuring
  temperature,'' \emph{IEEE Sensors Journal}, vol.~15, no.~8, pp. 4496--4507,
  2015.

\bibitem{5595724}
C.~Luo, F.~Wu, J.~Sun, and C.~W. Chen, ``Efficient measurement generation and
  pervasive sparsity for compressive data gathering,'' \emph{IEEE Transactions
  on Wireless Communications}, vol.~9, no.~12, pp. 3728--3738, 2010.

\bibitem{6168432}
W.~Chen and I.~Wassell, ``Energy-efficient signal acquisition in wireless
  sensor networks: a compressive sensing framework,'' \emph{IET Wireless Sensor
  Systems}, vol.~2, no.~1, pp. 1--8, 2012.

\bibitem{6548096}
G.~Yang, V.~Tan, C.~K. Ho, S.~H. Ting, and Y.~L. Guan, ``Wireless compressive
  sensing for energy harvesting sensor nodes,'' \emph{IEEE Transactions on
  Signal Processing}, vol.~61, no.~18, pp. 4491--4505, Sept 2013.

\bibitem{7390294}
W.~{Chen} and I.~J. {Wassell}, ``Cost-aware activity scheduling for compressive
  sleeping wireless sensor networks,'' \emph{IEEE Transactions on Signal
  Processing}, vol.~64, no.~9, pp. 2314--2323, 2016.

\bibitem{8480642}
P.~{Sun}, Z.~{Tian}, Z.~{Wang}, and Z.~{Wang}, ``Prss: A prejudiced random
  sensing strategy for energy-efficient information collection in the internet
  of things,'' \emph{IEEE Internet of Things Journal}, vol.~6, no.~2, pp.
  2717--2728, 2019.

\bibitem{baron2006distributed}
D.~Baron, M.~Wakin, M.~Duarte, S.~Sarvotham, and R.~Baraniuk, ``{Distributed
  compressed sensing},'' \emph{Technical Report ECE-0612, Electrical and
  Computer Engineering Department, Rice University}, Dec. 2006.

\bibitem{6502243}
M.~Duarte, M.~Wakin, D.~Baron, S.~Sarvotham, and R.~Baraniuk, ``Measurement
  bounds for sparse signal ensembles via graphical models,'' \emph{IEEE
  Transactions on Information Theory}, vol.~59, no.~7, pp. 4280--4289, 2013.

\bibitem{8532355}
J.~{Liu}, K.~{Huang}, and X.~{Yao}, ``Common-innovation subspace pursuit for
  distributed compressed sensing in wireless sensor networks,'' \emph{IEEE
  Sensors Journal}, vol.~19, no.~3, pp. 1091--1103, 2019.

\bibitem{6612900}
H.~Besbes, G.~Smart, D.~Buranapanichkit, C.~Kloukinas, and Y.~Andreopoulos,
  ``Analytic conditions for energy neutrality in uniformly-formed wireless
  sensor networks,'' \emph{IEEE Transactions on Wireless Communications},
  vol.~12, no.~10, pp. 4916--4931, 2013.

\bibitem{5948420}
M.~Mishali, Y.~Eldar, and A.~Elron, ``Xampling: Signal acquisition and
  processing in union of subspaces,'' \emph{IEEE Transactions on Signal
  Processing}, vol.~59, no.~10, pp. 4719--4734, 2011.

\bibitem{6155205}
F.~Chen, A.~Chandrakasan, and V.~Stojanovic, ``Design and analysis of a
  hardware-efficient compressed sensing architecture for data compression in
  wireless sensors,'' \emph{IEEE Journal of Solid-State Circuits}, vol.~47,
  no.~3, pp. 744--756, 2012.

\bibitem{5456165}
J.~Tropp and S.~Wright, ``Computational methods for sparse solution of linear
  inverse problems,'' \emph{Proceedings of the IEEE}, vol.~98, no.~6, pp.
  948--958, 2010.

\bibitem{1315936}
D.~P. {Wipf} and B.~D. {Rao}, ``Sparse bayesian learning for basis selection,''
  \emph{IEEE Transactions on Signal Processing}, vol.~52, no.~8, pp.
  2153--2164, 2004.

\bibitem{7558157}
W.~{Chen}, D.~{Wipf}, Y.~{Wang}, Y.~{Liu}, and I.~J. {Wassell}, ``Simultaneous
  bayesian sparse approximation with structured sparse models,'' \emph{IEEE
  Transactions on Signal Processing}, vol.~64, no.~23, pp. 6145--6159, 2016.

\bibitem{8453881}
W.~{Chen}, ``Simultaneously sparse and low-rank matrix reconstruction via
  nonconvex and nonseparable regularization,'' \emph{IEEE Transactions on
  Signal Processing}, vol.~66, no.~20, pp. 5313--5323, 2018.

\bibitem{8861380}
W.~{Chen}, X.~{Gong}, and N.~{Song}, ``Nonconvex robust low-rank tensor
  reconstruction via an empirical bayes method,'' \emph{IEEE Transactions on
  Signal Processing}, vol.~67, no.~22, pp. 5785--5797, 2019.

\bibitem{8332502}
X.~{Shen} and Y.~{Gu}, ``Nonconvex sparse logistic regression with weakly
  convex regularization,'' \emph{IEEE Transactions on Signal Processing},
  vol.~66, no.~12, pp. 3199--3211, 2018.

\bibitem{BAI2020107729}
\BIBentryALTinterwordspacing
Y.~Bai, W.~Chen, J.~Chen, and W.~Guo, ``Deep learning methods for solving
  linear inverse problems: Research directions and paradigms,'' \emph{Signal
  Processing}, p. 107729, 2020. [Online]. Available:
  \url{http://www.sciencedirect.com/science/article/pii/S0165168420302723}
\BIBentrySTDinterwordspacing

\bibitem{7500416}
G.~{Smart}, J.~{Atkinson}, J.~{Mitchell}, M.~{Rodrigues}, and
  Y.~{Andreopoulos}, ``Energy harvesting for the internet-of-things:
  Measurements and probability models,'' in \emph{2016 23rd International
  Conference on Telecommunications (ICT)}, 2016, pp. 1--6.

\bibitem{vullers2010energy}
R.~J. Vullers, R.~Schaijk, H.~J. Visser, J.~Penders, and C.~V. Hoof, ``Energy
  harvesting for autonomous wireless sensor networks,'' \emph{IEEE Solid-State
  Circuits Magazine}, vol.~2, no.~2, pp. 29--38, 2010.

\bibitem{pinuela2012current}
M.~Pinuela, D.~Yates, S.~Lucyszyn, and P.~Mitcheson, ``{Current state of
  research at Imperial College London in RF harvesting and inductive power
  transfer},'' in \emph{Proc. 2nd Int. Workshop Wireless Energy Transp.
  Harvesting, Leuven, Belgium, May}, 2012.

\bibitem{intellabWSN}
\BIBentryALTinterwordspacing
P.~Bodik, W.~Hong, C.~Guestrin, S.~Madden, M.~Paskin, and R.~Thibaux. (2004,
  Feb.) Intel lab data. [Online]. Available:
  \url{http://db.csail.mit.edu/labdata/labdata.html}
\BIBentrySTDinterwordspacing

\bibitem{deligiannis2014maximum}
N.~Deligiannis, A.~Munteanu, S.~Wang, S.~Cheng, and P.~Schelkens, ``{Maximum
  likelihood Laplacian correlation channel estimation in layered Wyner-Ziv
  Coding},'' \emph{IEEE Transactions on Signal Processing}, vol.~62, no.~4, pp.
  892--904, 2014.

\bibitem{roundy2004power}
S.~Roundy, D.~Steingart, L.~Frechette, P.~Wright, and J.~Rabaey, ``Power
  sources for wireless sensor networks,'' \emph{Wireless Sensor Networks}, pp.
  1--17, 2004.

\bibitem{varodayan2006rate}
D.~Varodayan, A.~Aaron, and B.~Girod, ``Rate-adaptive codes for distributed
  source coding,'' \emph{Signal Processing}, vol.~86, no.~11, pp. 3123--3130,
  2006.

\bibitem{verbist2013encoder}
F.~Verbist, N.~Deligiannis, S.~M. Satti, P.~Schelkens, and A.~Munteanu,
  ``Encoder-driven rate control and mode decision for distributed video
  coding,'' \emph{EURASIP Journal on Advances in Signal Processing}, vol. 2013,
  no.~1, p. 156, 2013.

\bibitem{1328091}
Z.~Xiong, A.~Liveris, and S.~Cheng, ``Distributed source coding for sensor
  networks,'' \emph{IEEE Signal Processing Magazine}, vol.~21, no.~5, pp.
  80--94, 2004.

\bibitem{akkouchi2008convolution}
M.~Akkouchi, ``On the convolution of exponential distributions,'' \emph{Journal
  of the Chungcheong Mathematical Society}, vol.~21, no.~4, pp. 501--510, 2008.

\end{thebibliography}

\end{document}